\documentstyle[aps,prl,preprint]{revtex}
\begin{document}
\title{The Anomalous Hall Effect in YBa$_2$Cu$_3$O$_7$}
\author{Branko P.\ Stojkovi\'c and David Pines}
\address{Department of Physics and Materials Research
Laboratory,\\
1110 West Green Street, University of Illinois, Urbana, IL 61801 }
\date{\today}
\maketitle
\begin{abstract}
The temperature dependence of the normal state Hall effect and
magnetoresistance in YBa$_2$Cu$_3$O$_7$ is investigated using
the Nearly Antiferromagnetic Fermi Liquid description of planar
quasiparticles. We find that highly anisotropic
scattering at different regions of the Fermi surface gives rise to
the measured anomalous temperature dependence of the resistivity and
Hall coefficient while yielding the universal temperature
dependence of the Hall angle observed
for both clean and dirty samples. This universality
is shown to arise from the limited momentum transfers available
for the anomalous, spin fluctuation scattering and is preserved for any
system with strong antiferromagnetic correlations.
\end{abstract}


\narrowtext
Eight years of experiments on increasingly pure samples of YBa$_2$Cu$_3$O$_7$
(YBCO) have demonstrated that, while ARPES
experiments suggest that the planar quasiparticles possess a
well defined Fermi Surface (FS), none of the
spin\cite{charlie} and charge\cite{donbook}
properties are those of a Landau Fermi Liquid (FL).
NMR experiments show that there are strong antiferromagnetic (AF) correlations
between neighboring Cu$^{2+}$
spins and that the spin-spin correlation function is strongly peaked at
${\bf Q} = (\pi,\pi)$, while
transport measurements yield a planar resistivity which is approximately linear
in temperature. The resistivity and optical properties are consistent with
a quasiparticle spectrum for which the Im$\Sigma({\bf p},\omega)$
is linear in the max($\omega,T$), for {\bf p} close to the FS, while the
effective mass enhancement is believed to be about 2. One of us has suggested
that the above anomalies can be traced to a strong magnetic
interaction between the quasiparticles\cite{dpreview}.
The resulting system is called a Nearly Antiferromagnetic
Fermi Liquid (NAFL).

While the NAFL model has been shown to be consistent, both
qualitatively and quantitatively, with
many experiments, including NMR, optical conductivity, resistivity,
the superconducting
transition temperature, and the role played by impurity
scattering\cite{dpreview},
a major challenge for the NAFL approach has been explaining
anomalous
transport in an applied magnetic field, where
experiments show that the Hall conductivity is a strong function of
temperature, yet the cotangent of the Hall angle has a very simple,
universal behavior
${\rm cot} \Theta_H = A + B T^2$. This universality has been explained in terms
of spin-charge separation\cite{pwa} and has been considered
as major support for that approach.

In this letter we report on calculations based on the NAFL model, of
the Hall conductivity $\sigma_H$
and resistivity $\rho$, using standard Boltzman transport theory.
Recently Hlubina and Rice (HR)\cite{tmrice}
explored this approach by
solving the Boltzman Equation (BE) using a variational method\cite{ziman}.
For $\rho(T)$
they found qualitative agreement
with the earlier self-consistent Eliashberg
calculation of Monthoux and Pines (MP)\cite{MP}.
Here we extend the work of HR by solving the BE
in a finite magnetic field $B$.
We use the same band parameters as HR and MP, $t=0.25$eV and
$t'=-0.45t$ with $\mu=-1.465t$, in agreement with ARPES.
We assume that the effective ({\em magnetic})
interaction between quasiparticles is given by\cite{MBP,MMP}
\begin{equation}
V^{eff}({\bf q}) = g^2\chi({\bf q},\omega) = {g^2\,A\over \omega_{sf} +
\omega_{sf} \xi^2 ({\bf q - Q})^2 - i\omega}\label{eq:xi}
\end{equation}
where the spin-fluctuation (SF) energy
$\omega_{sf}=T_0 + \beta\, T$, and $\xi$ is the magnetic correlation length
in units of the lattice constant $a$.
For optimally doped YBCO
the dimensionless parameter $A\approx 1.1$,
$T_0\approx 110$K and $\beta=0.55$,
and the energy $\omega_{sf}\xi^2 \approx 880$K is
very nearly independent of temperature\cite{MP}.

In Boltzman theory one seeks the displacement $\delta f_{\bf k} = f_{\bf k}
- f_0 = - \Phi_{\bf k} (\partial f_0/\partial \epsilon)$
of the Fermi Surface (FS) in an external field, where $f_0$ is the equilibrium
Fermi distribution function. 
The transport coefficients are found straightforwardly
from ${\bf j} = e \sum_{\bf k} {\bf v}\, \delta f_{\bf k}$.
We introduce the dimensionless quantities:
${\bf u}\equiv {\bf v} \hbar / at$, $b = B/B_0$,
where $B_0=\hbar c/e a^2\approx 4300T$.
Obviously, $b\ll 1$ for all applied fields of interest. 
We assume that ${\bf E} = E\hat{\bf x}$ and ${\bf B} \parallel
\hat{\bf z}$.
Then, the BE can be written as:
\begin{equation}
\Phi_{\bf k} = \left[u_x - b\,
\left(u_y {\partial\Phi_{\bf k}\over \partial k_x}
- u_x {\partial\Phi_{\bf k}\over \partial k_y}\right)\right]\,f_0 ({\bf k})
/I({\bf k},T)
\label{eq:be}
\end{equation}
where the linearized collision term is given by
\begin{equation}
I({\bf k},T) = {g^2\over t^2}
\int d\epsilon'\int {dk'\over 2\pi^2\vert {\bf u}\vert}
n(\epsilon_{\bf k} - \epsilon_{\bf k'})
\,{\rm Im}\,\chi ({\bf k-k'},\epsilon_{\bf k} - \epsilon_{\bf k'})
\left({\Phi_{\bf k'}\over \Phi_{\bf k}} - 1\right) f_0 ({\bf k'})
\label{eq:i0}
\end{equation} 
We solve the non-linear integral Eqs.\ (\ref{eq:be})
and (\ref{eq:i0})
numerically
on a fine mesh in momentum
space with typically 200$\times$200 points in the Brillouin zone (BZ). 
We have verified that our results for the Hall conductivity practically
do not vary for mesh sizes larger than 100$\times$100. 
Our principal results are given in Figs.\ \ref{fig:rho}-\ref{fig:cot}.
The input parameters in the figures are as specified above
for the fully doped YBCO, unless stated otherwise.

As may be seen in Fig.\ \ref{fig:rho} our calculated result for the
resistivity as a function of temperature
is qualitatively similar to HR and MP;
quantitatively it is closer to MP, especially at lower temperatures.
This is as expected, since MP included lifetime effects
by self-consistently calculating the real part of the
self-energy $\Sigma$. 
HR found a somewhat higher resistivity,
possibly due to their choice of variational function. It is
encouraging that our result is close to the experimentally observed values.
Note, however, that both HR and
the present calculation show a small non-linearity in $\rho(T)$,
due to normal FL--like scattering. The latter  contribution
disappears at increased temperature, as can be seen in the inset of
Fig.\ \ref{fig:rho}. We return to this point later.

Our results for the
Hall conductivity (Fig.\ \ref{fig:sigmaxy})
are in good
qualitative agreement with the experimental
results of Ref.\ \cite{gins90},
although they overestimate the size of $\sigma_H$.
The inset, which shows the same quantity as a function of magnetic
field, indicates that we are in a low field regime, as is expected
for a system with strong scattering\cite{gins91,hurd}.
Figure \ref{fig:cot} is our main result: it shows that
$\cot \Theta_H$ obeys the simple form $A+BT^2$, in agreement with
experiment.

We now consider the physical origin
of these results, beginning with the resistivity.
At first sight the above results seem surprising, since in a FL
the temperature behavior
of the Hall angle is the same as that of the
resistivity and one assumes a uniform
temperature dependence of scattering along the FS.
However, this is not the case for a NAFL where the
effective interaction (\ref{eq:xi}) depends
strongly on the momentum transfer ${\bf q}$. As the calculations
of HR and MP show, the quasiparticle lifetime $\tau_{\bf k}$
is highly anisotropic, because the scattering is very
strong for points ${\bf k}$ and ${\bf k'}$ such that
${\bf k-k'}={\bf q}\approx {\bf Q}$. As may be seen in Fig.\ \ref{fig:hot},
there are only a small
number of such {\em regions\/} on the FS, called ``hot spots'' by HR.
Nevertheless, as HR have shown\cite{tmrice},
for $\epsilon_{\bf k}\gg T_0$
the {\em average\/} quasiparticle lifetime is linear in the excitation
energy $\epsilon_{\bf k}$. In addition, the {\bf k} dependence of $\tau$
is found to be large only away from hot spots on the Fermi surface,
in regions where the scattering is no longer anomalous.
It is the combination of these features which is responsible for the
peculiar temperature dependence of both resistivity and the Hall angle.

For our orientation of ${\bf E}$, with $B=0$,
$\Phi_{\bf k}(B=0)\equiv \Phi_{\bf k}^0$
can be written
as $u_x/I_0({\bf k},T)$ where $I_0$ is given by Eqs.\ (\ref{eq:be})
and (\ref{eq:i0}), while in a
weak field $B$, one has $\Phi(B) \approx \Phi^0 -
(b/I_0) [u_y\,(\partial \Phi^0 / \partial k_x) - u_x\,(\partial \Phi^0 /
\partial k_y)]$. The conductivity
$\sigma_0$ and the Hall conductivity
$\sigma_H$ are averages of $u_x\Phi_{\bf k}$ and
$u_y\Phi_{\bf k}$ around the FS; since
$I_0$ possesses the symmetry of the crystal lattice,
the contribution of $\Phi_0$ to $\sigma_H$ vanishes
identically. Thus, $\sigma_0$ is an overage
of $u_x^2/I_0$ and $\sigma_H$ is an average of
$u_y\,(b/I_0)\, [u_y\,\nabla_x(u_x/I_0) - u_x\,\nabla_y(u_x/I_0)]$
around the FS.
The integral over $\epsilon'$ in Eq.\ (\ref{eq:i0})
can be done analytically for ${\bf k}$ on the FS. It
equals $\pi T\,[(2\omega_{k-k'})^{-1} + \sum_n (-1)^n \,
(\omega_{k-k'} + n\pi T)^{-1}]$, where $\omega_{q}= \omega_{sf} +
\omega_{sf} \xi^2 ({\bf q - Q})^2$, and can be
approximated by $\pi T^2/2\omega_{\bf k-k'}(\omega_{\bf k-k'}+\pi T)$.
When $T_0 \ll T \ll \omega_{sf} \xi^2$ we find, for ${\bf k}$ in
the middle of a hot spot, that $\Phi^0_{\bf k}$ is independent of ${\bf k}$,
$\Phi^0 \sim u_x/\sqrt{T/\omega_{sf}\xi^2}$. $\Phi^0_{\bf k}$ becomes
${\bf k}$ dependent away from a hot spot
and is given by $\Phi^0_{\bf k}\sim u_x\,s^3/(T/\omega_{sf}\xi^2)^2$,
where $s$ is the distance along the FS between ${\bf k}$
and the center of the
nearest hot spot. Experiment shows that for YBCO $\xi\sim 2$
in temperature region $100$K$<T<300$K (Ref.\ \cite{BP}).
This means that the hot
regions are fairly large:
under these circumstances, ${\bf k}$
is never too far from a hot spot (see Fig.\ \ref{fig:hot}).
Since $\Phi_{\bf k}$ must have
a maximum somewhere in-between two hot spots along the FS,
the $s$ dependence of $\Phi$ is significantly altered:
for $s\sim \sqrt{T/\omega_{sf}}/\xi$ we
now have $\Phi\sim s/(T/\omega_{sf}\xi^2)^2$
over most of the FS.
As a result $\sigma\sim 1/T$,
and  $\rho\sim g^2 (T/\omega_{sf}\xi^2)$. As shown in the inset of
Fig.\ \ref{fig:rho}, a 10\% change in $\omega_{sf}\xi^2$
yields a change of slope in resistivity of about 10\%.
Note that so far we have
only included the fact that $\xi\sim 1$ and that the hot
regions are symmetrically placed on the FS.
It is important to note that the contribution of the equidistant
regions from two adjacent hot spots is still $\sim (T/\omega_{sf}\xi^2)^2$.
This is why one finds
a small non-linearity in $\rho(T)$ at lower temperatures,
which vanishes as $T$ becomes larger than $T_0$ (see Fig.\ \ref{fig:rho},
and its inset) and the hot spots spread as $s\sim \sqrt{T}$.
Our result is also consistent with the anisotropy
of the self-energy found earlier\cite{MP}.

The above arguments are independent of any special features of the FS.
The actual shape of the FS is of little relevance to $\sigma_0$,
provided it is large enough
to allow for spin fluctuation scattering at ${\bf Q}$ (Ref.\ \cite{MP})
and the Fermi velocity is finite everywhere on the FS.
However, the Hall effect does depend quantitatively
on the detailed shape of the FS. In general, the FS of several HTSC families
is rather flat near hot spots, and curved away from them. Since $\sigma_H$
involves a gradient of $\Phi$, it is plausible that in the
regions where $I_0$ varies rapidly with $s$,
this gradient is dominated by
the gradient of $I_0$; moreover,
the curved regions of the FS will contribute to the
Hall effect mostly through the change in $u_x$.
One easily finds that in the flat regions
there is an additional temperature
contribution, $\partial \Phi/
\partial s\sim (T/\omega_{sf}\xi^2)^2$, to the Hall conductivity
as opposed to resistivity, while in the curved regions
one finds an additional contribution, $\nabla u_x/(T/\omega_{sf}\xi^2)^2$.
Therefore, in both regions the Hall conductivity has an extra
factor of $(T/\omega_{sf}\xi^2)^{-2}$, which should show clearly in
the temperature dependence of
the Hall angle, regardless of the presence of
the non-linear terms in the resistivity.
This is indeed the case, as clearly shown in the inset of Fig.\ \ref{fig:cot},
where we have plotted $\cot \Theta_H$ for two values of $\xi$.
Notice a very slight bending of the two lines at very high temperatures.
The present theory assumes that $T\ll \omega_{sf}\xi^2$; when this
is no longer the case, FL theory breaks down and
$\cot \Theta_H$ exhibits a deviation from the above universal
behavior. We have verified by explicit calculation that for fully doped
YBCO this occurs at approximately 700K (Ref.\ \cite{SPU}).

An important test of the soundness of our approach is obtained by adding
weak impurity scattering to the collision term in the BE.
Impurity scattering might be expected to add a temperature
independent term to $I_0$, so that $\cot \Theta_H$
should be proportional to $A+BT^2$, where the slope $B$
is unchanged from the clean case and $A$ is
proportional to the impurity concentration $n_i$ (Ref.\ \cite{SPU}).
The results of our numerical calculations  (Fig.\ \ref{fig:cot})
show that this is indeed the case:
$B$ is the same in clean and dirty cases to within 1\%,
while $A$ is proportional to $n_i$, for $n_i\sim 1$\%.
These results are
consistent with a number of experiments\cite{ong,forro,cooper}.

Another challenge for the NAFL model is the existence of a small
orbital magnetoresistance (MR), which, as has recently been observed in
YBCO by Harris {\em et al\/}\cite{ong-mr}, does not obey the
usual K\"ohler rule\cite{SPU}.
In calculations of $\rho(B)$ at
several temperatures we have verified this behavior.
Not to our surprise, the relative MR,
$\Delta\rho/\rho$, is found to have very strong temperature dependence
and is of order $\Theta_H^2$, as is usually the case in FLs.
The values we find are somewhat higher than those obtained
experimentally\cite{SPU}. 

In conclusion, we have shown that the energy and momentum dependence
of the effective magnetic interaction
in the NAFL model gives rise to an anomalous temperature dependence of
both resistivity and the Hall conductivity, which
are consistent with {\em all\/} transport measurements on fully
doped YBa$_2$Cu$_3$O$_7$.
The anomalous behavior originates
in the hot spots which have a region of influence, $\delta k \sim \sqrt{T}$:
as long as $\chi$ is sharply peaked at ${\bf q}
\approx {\bf Q}$, with $\xi\sim 1$,
this inevitably leads to the universal temperature behavior
of $\cot\Theta_H$.
Our results depend strongly on the
magnetic correlation length $\xi$, but are otherwise rather insensitive
to the band parameters\cite{SPU}. Their sensitivity to $\xi$
provides a natural explanation of the experimentally observed
noticeable differences in the slope of resistivity for rather
small variations in $T_c$ (Ref.\ \cite{donbook}).
In addition, since our results depend for the most part
on the symmetric
placement of the hot spots on the FS, and the existence of
strong AF correlations,
we expect similar results will hold for other HTSCs.

There are several unresolved issues. We always find
a somewhat lower resistivity with a somewhat higher slope
and a more negative intercept
than that measured
at lower temperatures. 
The slope depends strongly on the coupling constant $g$ and the
energy scale $\omega_{sf}\xi^2$, which are known only to
within 10\% accuracy; it may also be altered by strong coupling
corrections.
There are several possible explanations
for the the large negative intercept, which leads to
somewhat higher
values of both $\sigma_H$ and the MR than are seen experimentally:
small impurity scattering, spin-gap
effects, and normal Fermion scattering, all seen
experimentally in NMR measurements\cite{charlie,MBP}.
Imperfections which alter the local magnetic order
change the magnetic quasiparticle interaction\cite{MP}, and therefore
affect the resistivity,  while
shifting $\cot\Theta_H$ only marginally. Indeed, only the highest
quality samples show small negative intercepts in $\rho(T)$\cite{gins-i}.
Our calculations
show that an impurity concentration of 1\% reduces the
size of $\sigma_H$ to slightly below the experimentally obtained values,
without significantly altering its qualitative behavior,
while in addition to providing a residual resistivity,
impurities act to
remove the non-linearity seen in Fig.\ \ref{fig:rho} (Ref.\ \cite{SPU}).
The spin pseudogap effect produces
a curvature of the resistivity slightly above $T_c$ (Ref.\ \cite{BP}),
while normal Fermion scattering\cite{MBP}
reduces the Hall conductivity and somewhat
increases the resistivity. We address these issues elsewhere\cite{SPU}.

We are indebted to V.\ Barzykin, G.\ Blumberg,
D.\ Ginsberg, A.\ J.\ Leggett, M.\ Lercher,
N.\ P.\ Ong, T.\ M.\ Rice, Q.\ Si, C.\ P.\ Slichter, A.\ Sokol,
M.\ B.\ Weissman and Y.\ Zha for useful conversations.
We thank the National Center for
Supercomputing Applications for a grant of computer time.
This research is supported in part by NSF through
grants NSF-DMR 89-20538 (MRL at UIUC) and NSF-DMR 91-20000 (STCS).

\begin{figure}
\caption{The resistivity as a function of temperature in a NAFL (solid line).
The dotted line shows the result of HR [5],
the open circles show the result of MP [7], and triangles the
experimentally obtained $\rho_{aa}$ (Ref.\ [10]).
We plot $\rho$ only for $T>125$K, since
a {\em spin gap}, which may
alter the interaction (1), forms below this temperature [14].
The parameters are given in the text. Inset: the NAFL results for
$\omega_{sf}\xi^2=880$K (solid line) and 970K (dashed line)
at higher temperatures.}
\label{fig:rho}
\end{figure}

\begin{figure}
\caption{The Hall conductivity $\sigma_H$
as a function of temperature. The solid line
shows the calculated result, while the triangles show the
experimental results of Ref.\ [10]. Inset: NAFL results for $\sigma_H$
as a function of the applied magnetic field at (top to bottom)
$T=150$, 200 and 250K.}
\label{fig:sigmaxy}
\end{figure}

\begin{figure}
\caption{Cotangent of the Hall angle as a function of temperature for
a NAFL. The solid and dotted lines represent
clean and dirty
samples respectively. Inset: results at
higher $T$ for $\omega_{sf}\xi^2=880$K (solid line) and 970K (dashed line).}
\label{fig:cot}
\end{figure}

\begin{figure}
\caption{The FS of YBCO (solid line)
and the magnetic BZ (dashed line). The hot spots are
the thick regions near the intercepts of the two lines.}
\label{fig:hot}
\end{figure}

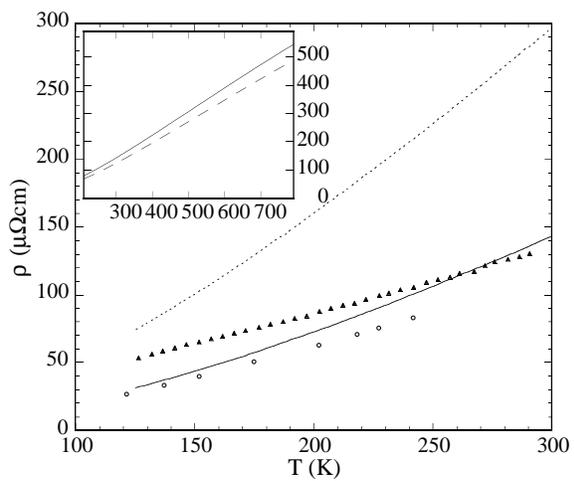
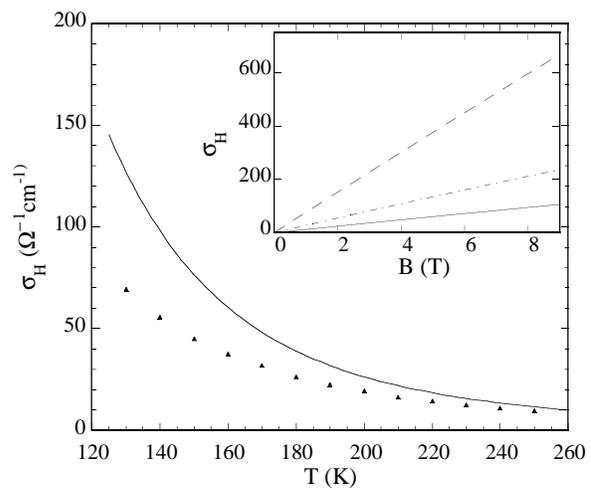
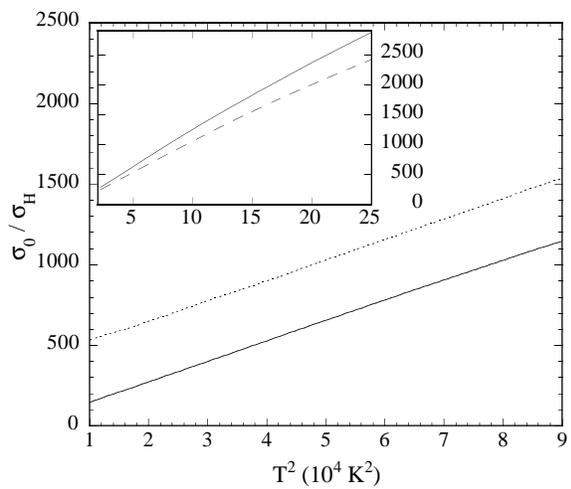
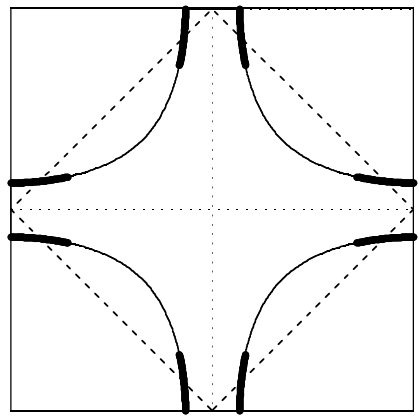
\end{document}